\documentclass[a4paper,fleqn]{cas-dc}
\usepackage{xcolor}
\colorlet{scolor}{black}

\usepackage[numbers]{natbib}
\usepackage{tikz}
\usepackage{algorithm}
\usepackage{subcaption}
\usepackage{algpseudocode} 
\usetikzlibrary{patterns}
\usetikzlibrary{arrows.meta}
\def\tsc#1{\csdef{#1}{\textsc{\lowercase{#1}}\xspace}}
\tsc{WGM}
\tsc{QE}
\tsc{EP}
\tsc{PMS}
\tsc{BEC}
\tsc{DE}

\begin{document}
\let\WriteBookmarks\relax
\def\floatpagepagefraction{1}
\def\textpagefraction{.001}

\shorttitle{}

\shortauthors{A.F. Bolaños-Acosta \& J.C. Restrepo.}

\title [mode = title]{On the Assessment of Domain Decomposition Approaches in Extended Physics-Informed Neural Networks for the Shock-Dominated Burgers Equation}                      

%

%
\author[1]{{Andrés F. Bolaños-Acosta}}[orcid=0000-0002-4846-0861]


\fnmark[]

\ead{afb.acosta@gmail.com}


\credit{Writing - original draft, Methodology, Investigation, Formal analysis, Conceptualization}

\affiliation[1]{organization={Independent Researcher},
    city={Pasto},
    country={Colombia}}

\author[2]{{Julián C. Restrepo}}[orcid=0000-0003-3519-5744]
\affiliation[2]{organization={FLUX - Thermal systems Lab, Department of Aeronautical Engineering, School of Engineering of São João da Boa Vista, Universidade Estadual Paulista},
	city={São Paulo},
	postcode={13876–750}, 
	country={Brazil}}
\cormark[1]

\ead{julian.restrepo@unesp.br}

\credit{Writing - review \& editing, Supervision, Resources, Project administration}






\begin{abstract}
Physics-Informed Neural Networks (PINNs) provide a mesh-free framework for solving partial differential equations but often struggle with sharp gradients and shock-dominated solutions. Extended Physics-Informed Neural Networks (XPINNs) address this challenge through domain decomposition; however, the effects of decomposition design, overlap, and local network capacity on solution accuracy remain insufficiently characterized. This work presents a systematic assessment of domain decomposition configurations for XPINNs applied to the one-dimensional viscous Burgers equation ($\nu=0.01/\pi$). We compare a standard PINN, a conventional XPINN, a shock-focused XPINN with increased network capacity in the shock-containing subdomain, and overlapping XPINNs with different overlap widths. All configurations are trained using a common protocol based on Adam, L-BFGS refinement, and ten Residual-based Adaptive Refinement (RAR) cycles. Performance is evaluated against a reference solution using global and shock-region relative $L_2$ and $L_\infty$ errors, maximum PDE residuals, and interface continuity errors. The results show that decomposition design strongly affects XPINN accuracy and shock resolution. Among the overlapping configurations, the smallest overlap provides the best accuracy, while larger overlaps degrade performance. The shock-focused XPINN achieves the lowest overall solution errors, indicating that increased local network capacity can provide greater benefit than increasing overlap alone. RAR produces the largest improvements during early refinement stages, followed by increasingly non-monotonic changes in accuracy. These findings highlight the importance of jointly considering decomposition, interface coupling, network capacity, and adaptive refinement for shock-dominated problems.
\end{abstract}



\begin{keywords}
Physics-Informed Neural Networks \sep
XPINNs \sep
Domain Decomposition \sep
Burgers Equation \sep
Shock-Dominated Flows
\end{keywords}

\maketitle

\section{Introduction}

Partial differential equations (PDEs) have long been a subject of interest in the scientific community because they describe complex physical phenomena that underpin engineering applications in various fields such as fluid mechanics, heat transfer, chemical reactions and phase transitions \cite{Singh2025}.  PDEs characterize these phenomena through mathematical models constrained by initial and boundary conditions, making many problems difficult or even impossible to solve analytically. To address these challenges, researchers have developed analytical solutions, numerical methods,  particularly within Computational Fluid Dynamics (CFD), and, more recently, Scientific Machine Learning (SciML) approaches \cite{LeVeque2002, Karniadakis2021}. 

Among the various PDEs used to evaluate numerical methods, the Burgers equation has become a standard benchmark because it represents a simplified form of the Navier--Stokes equations \cite{Burgers1974,LeVeque2002}. Under low-viscosity conditions, it reproduces key nonlinear phenomena, including shock-wave formation, propagation, and dissipation, while providing a controlled environment in which steep gradients naturally arise. Consequently, accurately resolving these shocks remains a significant challenge for both analytical and numerical approaches \cite{Raissi2019, Jagtap2020cpinns}.

Traditional numerical techniques, including the Finite Difference Method (FDM), Finite Element Method (FEM), and Finite Volume Method (FVM), have been extensively employed to solve PDEs involving steep gradients and discontinuities. However, accurately resolving these features often requires specialized numerical treatments to stabilize the solution while maintaining high accuracy in smooth regions \cite{Luo2025}. As a result, the performance of these methods strongly depends on mesh resolution and the adopted stabilization strategy. Although these techniques have demonstrated remarkable accuracy and robustness, their application to complex or multiscale problems often requires increasingly refined meshes, leading to increased computational cost \cite{Moseley2023}.

With the rapid advances in Machine Learning (ML) over the past decade, Physics-Informed Neural Networks (PINNs), introduced by \citet{Raissi2019}, have emerged as a powerful framework for solving PDEs by incorporating the governing physical laws directly into the training process. Unlike traditional numerical methods, PINNs provide a mesh-free approach that can be applied to both forward and inverse problems. In forward problems, the governing equations together with the initial and boundary conditions are enforced through the loss function, whereas in inverse problems, unknown parameters of the governing equations can be inferred from sparse observational data \cite{Cuomo2022, Shukla2021}.

Despite their promising capabilities, PINNs exhibit several limitations compared with established numerical methods. Their theoretical convergence properties remain an active area of research, and the associated optimization problem often becomes increasingly challenging when the target solution contains steep gradients or discontinuities \cite{Wang2021, Cuomo2022}. In addition, PINNs tend to learn low-frequency solution components more rapidly than high-frequency components, a phenomenon known as spectral bias, which hinders the accurate representation of sharp features \cite{Rahaman2019, Tancik2020}. From a computational perspective, standard PINNs generally require retraining for each problem instance, making conventional numerical solvers more efficient for many forward simulations. Furthermore, as the complexity and size of the computational domain increase, maintaining solution accuracy typically requires larger neural network architectures together with a greater number of collocation points, leading to increased computational cost and scalability challenges \cite{Moseley2023, Luo2025}. 

To overcome the limitations of standard PINNs, domain decomposition has emerged as one of the most effective strategies for improving training efficiency, scalability, and solution accuracy \cite{Klawonn2024,Shukla2021,Moseley2023,Dolean2024}. Among these approaches, Extended Physics-Informed Neural Networks (XPINNs), proposed by \citet{Jagtap2020xpinns}, provide a flexible framework in which the computational domain is partitioned into multiple subdomains, each represented by an independent neural network. This formulation allows each subnetwork to employ different architectures and capacities according to the local solution complexity, while interface conditions ensure communication between adjacent subdomains. By decomposing the learning task into smaller local problems, XPINNs improve solution representation, facilitate parallel training, and offer greater flexibility than standard PINNs. In addition to domain decomposition, complementary techniques such as adaptive activation functions \cite{Jagtap2020acfuncs}, Fourier feature mappings \cite{Tancik2020, Hou2026}, and Residual-based Adaptive Refinement (RAR) \cite{Lu2021, Wu2023} have been proposed to improve optimization, mitigate spectral bias, and enhance adaptive collocation.

Within the domain decomposition framework, \citet{Jagtap2020cpinns} first introduced Conservative Physics--Informed Neural Networks (cPINNs) for nonlinear conservation laws by enforcing conservative flux continuity across subdomain interfaces. This concept was subsequently generalized through XPINNs to arbitrary space-time decompositions and a broader class of partial differential equations. Other notable developments include hp-VPINNs, which combine the variational formulation of PINNs with h- and p-refinement strategies \cite{Kharazmi2021}, and Finite Basis Physics-Informed Neural Networks (FBPINNs), where the computational domain is partitioned into overlapping subdomains combined through smooth window functions to improve scalability and alleviate spectral bias \cite{Moseley2023}. More recently, \citet{Dolean2024} extended the FBPINN framework to a multilevel architecture inspired by classical Schwarz domain decomposition methods. Beyond domain decomposition, several formulations have been proposed to improve the approximation of shock-dominated solutions, including weak-form PINNs \cite{Wang2025}, control-volume PINNs (cvPINNs) \cite{Patel2022}, relaxation-based neural networks (RelaxNN)\cite{Zhou2024}, Godunov-Riemann informed neural networks (GoRINNs) \cite{Siettos2025}, and characteristic-based formulations \cite{Chen2025, Han2025}, all of which incorporate conservation principles, entropy constraints, finite-volume formulations, or characteristic information to improve the representation of discontinuities \cite{Abbasi2025}.

Although previous studies have investigated domain decomposition, interface treatments, and overlapping formulations, much of the existing work has focused on scalability, convergence, or optimization efficiency. A systematic assessment of domain decomposition strategies within the XPINN framework for nonlinear conservation laws remains comparatively limited. In particular, the influence of different decomposition configurations, including overlapping interface regions, on the representation of shock-dominated solutions, interface flux continuity, and overall solution accuracy has not been systematically characterized.

To address this gap, this work presents a systematic comparative study of domain decomposition configurations for the viscous Burgers equation. Four configurations are considered: a standard PINN, a conventional XPINN, an XPINN with increased network capacity in the shock-containing subdomain, and an overlapping XPINN formulation inspired by classical Schwarz domain decomposition methods \cite{Smith1996,Klawonn2024}. These configurations are designed to isolate the effects of spatial decomposition, local network capacity, and overlap width on the approximation of shock-dominated solutions. All models are trained using a common optimization protocol based on the Adam optimizer \cite{Kingma2015}, L-BFGS refinement \cite{Liu1989}, and residual-based adaptive refinement (RAR). Their performance is evaluated using global and shock-region relative $L_2$ and $L_\infty$ errors, maximum PDE residuals, and interface continuity errors. The results show that the shock-focused XPINN with increased local network capacity achieves the lowest overall solution errors among the configurations considered. For the overlapping XPINNs, the smallest overlap provides the best accuracy, whereas larger overlap regions degrade performance, indicating that increasing local network capacity can be more beneficial than increasing overlap alone for shock-dominated solutions.

The remainder of this paper is organized as follows. Section~2 presents the governing equations and the viscous Burgers formulation. Section~3 introduces the PINN, XPINN, and overlapping XPINN formulations. Section~4 describes the numerical setup and training methodology. Section~5 presents the numerical results and comparative analysis, while Section~6 discusses the implications and limitations of the findings. Finally, Section~7 summarizes the main conclusions and outlines directions for future research.

\section{Governing Equations}\label{sec:governing_equations}

Let $\Omega=[x_L,x_R]\subset\mathbb{R}$ be the entire spatial computational domain and let $I=[0,T]$ denote the temporal domain. The corresponding space--time domain is given by $\Omega\times I$. Throughout this work, the spatial and temporal domains are defined as $\Omega=[-1,1]$ and $I=[0,1]$, respectively.

The one-dimensional viscous Burgers equation \cite{LeVeque2002} is adopted as the model problem considered in this work. Since the objective is to investigate domain decomposition strategies for nonlinear conservation laws, the governing equation is expressed in conservative form as

\begin{equation}
	\frac{\partial u}{\partial t}
	+
	\frac{\partial}{\partial x}
	\left(
	\frac{u^{2}}{2}
	\right)
	=
	\nu
	\frac{\partial^{2}u}{\partial x^{2}},
	\qquad
	(x,t)\in\Omega\times I,
	\label{eq:burgers}
\end{equation}
where $u(x,t)$ denotes the solution field, with $x\in\Omega$ and $t\in I$ representing the spatial and temporal coordinates, respectively. The parameter $\nu>0$ denotes the kinematic viscosity coefficient, which controls the diffusion strength. For all numerical experiments presented in this work, the viscosity is fixed as
\begin{equation}
	\nu=\frac{0.01}{\pi},
	\label{eq:viscosity}
\end{equation}
corresponding to the low-viscosity regime commonly adopted for the shock-dominated Burgers benchmark.
The nonlinear convective flux is defined as
\begin{equation}
	f(u)=\frac{u^{2}}{2},
	\label{eq:convective_flux}
\end{equation}
whereas the total conservative flux is given by
\begin{equation}
	F(u,u_x)=
	\frac{u^{2}}{2}
	-
	\nu\frac{\partial u}{\partial x}.
	\label{eq:total_flux}
\end{equation}
Accordingly, Eq.~(\ref{eq:burgers}) can be rewritten in divergence form as
\begin{equation}
	\frac{\partial u}{\partial t}
	+
	\frac{\partial F(u,u_x)}{\partial x}
	=
	0,
	\label{eq:burgers_divergence}
\end{equation}
where the nonlinear convective flux $f(u)$ drives the formation and propagation of steep gradients and shock-like structures, while the diffusive contribution $-\nu\,\partial u/\partial x$ regularizes the solution by smoothing these gradients. The conservative formulation adopted in this work explicitly represents the total physical flux and provides the mathematical foundation for enforcing conservative flux continuity across subdomain interfaces \cite{Jagtap2020cpinns}, which forms the basis of the domain decomposition framework developed in this work.

The Burgers equation given by Eq.~(\ref{eq:burgers}) is supplemented with the following initial and boundary conditions:
\begin{align}
	u(x,0)
	&=
	-\sin(\pi x),
	&& x\in\Omega,
	\label{eq:ic}
	\\
	u(-1,t)
	&=
	0,
	&& t\in I,
	\label{eq:bc_left}
	\\
	u(1,t)
	&=
	0,
	&& t\in I.
	\label{eq:bc_right}
\end{align}

The sinusoidal initial condition evolves into a shock-dominated solution due to the nonlinear convective dynamics, while diffusion regularizes the discontinuity according to the prescribed viscosity coefficient. Homogeneous Dirichlet boundary conditions are imposed at both ends of the computational domain throughout the simulation.

\section{Physics-Informed Neural Network Formulation}
\label{sec:pinn}

\subsection{Standard PINN formulation}

Physics-Informed Neural Networks (PINNs) approximate the solution of a partial differential equation by means of a neural network while embedding the governing physical laws directly into the optimization process. Let $u_{\theta}(x,t)$ denote a neural network parameterized by the trainable parameters $\theta$, which provides the approximation
\begin{equation}
	u(x,t)\approx u_{\theta}(x,t),
	\label{eq:pinn_approximation}
\end{equation}

where $\theta$ represents the collection of all trainable weights and biases of the network. Unlike conventional supervised learning, the network is trained without requiring labeled solution data. Instead, the governing equation, together with the initial and boundary conditions, is enforced through a physics-informed loss function evaluated at a set of collocation points distributed throughout the computational domain.

Since the Burgers equation is considered in conservative form, the corresponding physics residual is defined as

\begin{equation}
	R_{\theta}(x,t)
	=
	\frac{\partial u_{\theta}}{\partial t}
	+
	\frac{\partial}{\partial x}
	F\!\left(
	u_{\theta},
	\frac{\partial u_{\theta}}{\partial x}
	\right),
	\qquad
	(x,t)\in\Omega\times I,
	\label{eq:pde_residual}
\end{equation}

where $F(u_{\theta},\partial u_{\theta}/\partial x)$ denotes the total conservative flux defined in Eq.~(\ref{eq:total_flux}). The required spatial and temporal derivatives are computed by automatic differentiation, allowing the governing equation to be enforced without constructing a numerical mesh.

The PDE residual loss is computed as the mean squared residual evaluated at the interior collocation points,
\begin{equation}
	\mathcal{L}_{\mathrm{PDE}}
	=
	\frac{1}{N_f}
	\sum_{i=1}^{N_f}
	R_{\theta}(x_i,t_i)^2,
	\label{eq:pde_loss}
\end{equation}
where $N_f$ denotes the number of interior collocation points.

Similarly, the initial condition is enforced through
\begin{equation}
	\mathcal{L}_{\mathrm{IC}}
	=
	\frac{1}{N_{\mathrm{IC}}}
	\sum_{i=1}^{N_{\mathrm{IC}}}
	\left(
	u_{\theta}(x_i,0)
	-
	u_0(x_i)
	\right)^2,
	\label{eq:ic_loss}
\end{equation}

where $u_0(x)$ denotes the prescribed initial condition.

The boundary conditions are imposed by minimizing
\begin{equation}
	\mathcal{L}_{\mathrm{BC}}
	=
	\frac{1}{N_{\mathrm{BC}}}
	\sum_{i=1}^{N_{\mathrm{BC}}}
	\left(
	u_{\theta}(x_i,t_i)
	-
	u_b(x_i,t_i)
	\right)^2,
	\label{eq:bc_loss}
\end{equation}

where $u_b$ represents the prescribed boundary values and $(x_i,t_i)$ are boundary collocation points.

The overall objective function is obtained as the weighted sum of the individual loss contributions,

\begin{equation}
	\mathcal{L}_{\mathrm{PINN}}
	=
	w_{\mathrm{PDE}}\mathcal{L}_{\mathrm{PDE}}
	+
	w_{\mathrm{ICBC}}
	\left(
	\mathcal{L}_{\mathrm{IC}}
	+
	\mathcal{L}_{\mathrm{BC}}
	\right).
	\label{eq:total_loss_pinn}
\end{equation}

where $w_{\mathrm{PDE}}$ and $w_{\mathrm{ICBC}}$ are the weighting coefficients associated with the governing-equation residual and the initial- and boundary-condition losses, respectively.

\subsection{Extended Physics-Informed Neural Networks}

Following \citet{Jagtap2020xpinns}, the computational domain is partitioned into $N_s$ non-overlapping subdomains,
\begin{equation}
	\Omega=\bigcup_{k=1}^{N_s}\Omega_k,
	\qquad
	\Omega_i\cap\Omega_j=\emptyset,
	\quad
	i\neq j,
	\label{eq:subdomains}
\end{equation}

where each subdomain $\Omega_k$ is assigned an independent neural network
$u_{\theta_k}(x,t)$ with trainable parameters $\theta_k$.

Accordingly, the global solution is approximated by the collection
\begin{equation}
	u(x,t)\approx
	\left\{
	u_{\theta_1},
	u_{\theta_2},
	\ldots,
	u_{\theta_{N_s}}
	\right\},
	\label{eq:xpinn_solution}
\end{equation}

where each $u_{\theta_k}$ provides a local approximation over its corresponding subdomain $\Omega_k$.

For each subdomain, the local physics residual is defined as
\begin{equation}
	R_{\theta_k}
	=
	\frac{\partial u_{\theta_k}}{\partial t}
	+
	\frac{\partial}{\partial x}
	F\!\left(
	u_{\theta_k},
	\frac{\partial u_{\theta_k}}{\partial x}
	\right),
	\qquad
	(x,t)\in\Omega_k.
	\label{eq:local_residual}
\end{equation}

Each subnetwork minimizes an independent loss function composed of PDE, initial-condition, and boundary-condition contributions,

\begin{equation}
	\mathcal{L}_k
	=
	w_{\mathrm{PDE}}
	\mathcal{L}_{\mathrm{PDE}}^{(k)}
	+
	w_{\mathrm{ICBC}}
	\left(
	\mathcal{L}_{\mathrm{IC}}^{(k)}
	+
	\mathcal{L}_{\mathrm{BC}}^{(k)}
	\right).
	\label{eq:local_loss}
\end{equation}

Unlike standard PINNs, XPINNs additionally require interface conditions to ensure consistency between neighboring subnetworks. These conditions enforce the continuity of the solution and, when appropriate, the conservative flux across subdomain interfaces. The interface formulation adopted in this work is presented in the following subsection.

\subsection{Conservative Interface Conditions}
Let $\Gamma_{ij}$ denote the interface separating two adjacent subdomains
$\Omega_i$ and $\Omega_j$, such that
\begin{equation}
	\Gamma_{ij}
	=
	\partial\Omega_i
	\cap
	\partial\Omega_j,
	\label{eq:interface}
\end{equation}

where $\partial\Omega_i$ and $\partial\Omega_j$ denote the boundaries of the corresponding subdomains.
To ensure a globally consistent approximation, the solution is required to be continuous across each interface. Accordingly, the following continuity condition is imposed,
\begin{equation}
	u_{\theta_i}
	=
	u_{\theta_j},
	\qquad
	\text{on } \Gamma_{ij},
	\label{eq:u_interface}
\end{equation}
where $u_{\theta_i}$ and $u_{\theta_j}$ denote the neural network approximations associated with the neighboring subdomains.
Since the Burgers equation is formulated in conservative form, continuity of the total conservative flux is additionally enforced across each interface. The corresponding condition is given by
\begin{equation}
	F_i
	=
	F_j,
	\qquad
	\text{on } \Gamma_{ij},
	\label{eq:flux_interface}
\end{equation}
where
\begin{equation}
	F_i
	=
	\frac{u_{\theta_i}^2}{2}
	-
	\nu
	\frac{\partial u_{\theta_i}}{\partial x},
	\qquad
	F_j
	=
	\frac{u_{\theta_j}^2}{2}
	-
	\nu
	\frac{\partial u_{\theta_j}}{\partial x}.
	\label{eq:interface_flux}
\end{equation}
The solution continuity loss is defined as
\begin{equation}
	\mathcal{L}_{u}^{\Gamma}
	=
	\frac{1}{N_\Gamma}
	\sum_{m=1}^{N_\Gamma}
	\left(
	u_{\theta_i}^{(m)}
	-
	u_{\theta_j}^{(m)}
	\right)^2,
	\label{eq:loss_interface_u}
\end{equation}
where $N_\Gamma$ denotes the number of interface collocation points.
Similarly, the conservative flux continuity loss is defined as
\begin{equation}
	\mathcal{L}_{F}^{\Gamma}
	=
	\frac{1}{N_\Gamma}
	\sum_{m=1}^{N_\Gamma}
	\left(
	F_i^{(m)}
	-
	F_j^{(m)}
	\right)^2.
	\label{eq:loss_interface_flux}
\end{equation}

The complete XPINN objective function is therefore written as
\begin{equation}
	\mathcal{L}_{\mathrm{XPINN}}
	=
	\sum_{k=1}^{N_s}
	\mathcal{L}_k
	+
	w_{\mathrm{interface}}
	\left(
	\mathcal{L}_{u}^{\Gamma}
	+
	\mathcal{L}_{F}^{\Gamma}
	\right)
	\label{eq:xpinn_loss}
\end{equation}
where $w_{\mathrm{interface}}$ is the weighting coefficient associated with the interface continuity constraints. The interface conditions described above correspond to the standard non-overlapping XPINN formulation. In the following subsection, these conditions are extended to overlapping subdomains by enforcing solution and conservative flux consistency throughout the overlap region rather than only at a single interface.

\subsection{Overlapping XPINN Formulation}
\label{sec:overlapping_xpinn}

The overlapping strategy considered in this work follows the general
principles of the classical Schwarz domain decomposition method
\cite{Smith1996}, in which neighboring subdomains share a finite region
to enhance information exchange. Accordingly, adjacent XPINN subdomains
are allowed to overlap, replacing the pointwise interface coupling of
the standard non-overlapping formulation by consistency constraints
defined over a finite spatial region.

Let $\Omega_i$ and $\Omega_j$ denote two neighboring subdomains. Their
overlap region is defined as
\begin{equation}
	\Omega_{ij}
	=
	\Omega_i\cap\Omega_j,
	\qquad
	\Omega_{ij}\neq\emptyset.
	\label{eq:overlap_region}
\end{equation}
For the one-dimensional spatial domain considered in this work, the
overlap is defined symmetrically around the nominal interface. Let
$x^\Gamma$ denote the corresponding interface location and let
$\delta$ denote the overlap extension from the nominal interface.
Then,
\begin{equation}
	\Omega_{ij}
	=
	[x^\Gamma-\delta,\,
	x^\Gamma+\delta],
	\label{eq:overlap_interval}
\end{equation}
and the resulting overlap width is
\begin{equation}
	|\Omega_{ij}|=2\delta.
	\label{eq:overlap_width}
\end{equation}

Within $\Omega_{ij}$, both subnetworks simultaneously approximate the
solution field,
\[
u_{\theta_i}(x,t),
\qquad
u_{\theta_j}(x,t),
\qquad
(x,t)\in\Omega_{ij}\times I,
\]
using independent trainable parameters. Consistency between neighboring
subnetworks is enforced by requiring agreement of both the solution and
the total conservative flux throughout the overlap region.

The overlap solution-continuity loss is defined as
\begin{equation}
	\mathcal{L}_{u}^{\Omega}
	=
	\frac{1}{N_o}
	\sum_{m=1}^{N_o}
	\left(
	u_{\theta_i}^{(m)}
	-
	u_{\theta_j}^{(m)}
	\right)^2,
	\label{eq:overlap_solution_loss}
\end{equation}
where $N_o$ denotes the number of collocation points sampled within the
overlap region.

The corresponding conservative flux-continuity loss is defined as
\begin{equation}
	\mathcal{L}_{F}^{\Omega}
	=
	\frac{1}{N_o}
	\sum_{m=1}^{N_o}
	\left(
	F_i^{(m)}
	-
	F_j^{(m)}
	\right)^2,
	\label{eq:overlap_flux_loss}
\end{equation}
where the conservative fluxes associated with the two subnetworks are
given by
\begin{equation}
	F_i
	=
	\frac{u_{\theta_i}^{2}}{2}
	-
	\nu
	\frac{\partial u_{\theta_i}}{\partial x},
	\qquad
	F_j
	=
	\frac{u_{\theta_j}^{2}}{2}
	-
	\nu
	\frac{\partial u_{\theta_j}}{\partial x}.
	\label{eq:overlap_flux}
\end{equation}

Unlike the standard XPINN formulation, in which neighboring subnetworks
are coupled through conditions imposed at a single interface, the
overlapping formulation enforces consistency at multiple shared
collocation points distributed throughout a finite spatial region.
This provides a broader coupling between neighboring subnetworks while
retaining the conservative flux constraint.

The resulting objective function for the overlapping XPINN formulation
is given by
\begin{equation}
	\mathcal{L}_{\mathrm{XPINN-OV}}
	=
	\sum_{k=1}^{N_s}
	\mathcal{L}_k
	+
	w_{\mathrm{overlap}}
	\left(
	\mathcal{L}_{u}^{\Omega}
	+
	\mathcal{L}_{F}^{\Omega}
	\right),
	\label{eq:xpinn_ov_loss}
\end{equation}
where $w_{\mathrm{overlap}}$ is the weighting coefficient associated
with the solution and conservative flux continuity constraints within
the overlap regions.

\section{Numerical Methodology}

\subsection{Network Architecture}

Each subnetwork introduced in the previous section is represented by an independent fully-connected feed-forward neural network. Given the space--time coordinates $(x,t)\in\Omega\times I$, the neural network defines a nonlinear mapping
\begin{equation}
	u_{\theta}(x,t)
	=
	\mathcal{N}_{\theta}(x,t),
	\label{eq:network_mapping}
\end{equation}

where $\theta$ denotes the set of trainable parameters composed of all weight matrices and bias vectors.

Let the input vector be defined as

\begin{equation}
	\mathbf{h}^{(0)}
	=
	\begin{bmatrix}
		x\\
		t
	\end{bmatrix}.
\end{equation}
The hidden layers are recursively computed according to
\begin{equation}
	\mathbf{h}^{(\ell)}
	=
	\phi
	\!\left(
	W^{(\ell)}
	\mathbf{h}^{(\ell-1)}
	+
	b^{(\ell)}
	\right),
	\qquad
	\ell=1,\ldots,L-1,
	\label{eq:hidden_layers}
\end{equation}

where $W^{(\ell)}$ and $b^{(\ell)}$ denote the weight matrix and bias vector associated with the $\ell$-th hidden layer, respectively, $\phi(\cdot)$ is the nonlinear activation function, and $L$ denotes the total number of network layers.

The network output is obtained through the final linear transformation
\begin{equation}
	u_{\theta}(x,t)
	=
	W^{(L)}
	\mathbf{h}^{(L-1)}
	+
	b^{(L)},
	\label{eq:network_output}
\end{equation}
which provides the approximation of the solution field over the corresponding subdomain. 

Since the governing equation is enforced through automatic differentiation, the neural network approximation remains fully differentiable with respect to its inputs. Consequently, all spatial and temporal derivatives required for evaluating the governing equation and the conservative flux are computed exactly through the computational graph without numerical discretization.

The specific activation function, parameter initialization strategy, network
depth, and width are specified in the experimental configuration section.

\subsection{Training Strategy}
The optimization problem is formulated as
\begin{equation}
	\theta^{*}
	=
	\arg\min_{\theta}
	\mathcal{L}(\theta),
	\label{eq:optimization_problem}
\end{equation}
where $\theta$ denotes the trainable network parameters and $\mathcal{L}$ corresponds to the loss function of the adopted formulation.

A hybrid Adam/L-BFGS optimization strategy is employed. Adam is first used for stochastic gradient-based optimization, followed by L-BFGS refinement to improve convergence.

To improve the resolution of localized features, the optimization is combined with RAR, which progressively enriches the collocation set by adding points associated with large physics residuals. The complete training procedure is summarized in Algorithm~\ref{alg:rar}.

\subsection{Residual-Based Adaptive Refinement}

RAR enriches the collocation set by identifying regions where the governing equation is poorly satisfied. Given a candidate set
\begin{equation}
	\mathcal{C}
	=
	\left\{
	(x_i,t_i)
	\right\}_{i=1}^{N_c},
	\label{eq:candidate_set}
\end{equation}
the residual magnitude is evaluated as
\begin{equation}
	r_i
	=
	\left|
	R_{\theta}(x_i,t_i)
	\right|,
	\qquad
	i=1,\ldots,N_c,
	\label{eq:rar_magnitude}
\end{equation}
Here, $R_{\theta}$ is the governing-equation residual defined in 
Section~\ref{sec:governing_equations}. Consequently, the residual 
magnitude $r_i$ provides a local measure of the degree to which the 
current neural-network approximation satisfies the governing equation 
at the candidate point $(x_i,t_i)$. Candidate points with larger values 
of $r_i$ therefore identify regions in which the current approximation 
exhibits a larger local physics error and are preferentially selected 
for refinement.

The $N_{\mathrm{c}}$ candidate points with the largest residuals are 
then added to the training set,
\begin{equation}
	\mathcal{D}^{(m+1)}
	=
	\mathcal{D}^{(m)}
	\cup
	\mathcal{S},
	\label{eq:rar_dataset}
\end{equation}
where $\mathcal{D}^{(m)}$ denotes the collocation-point set used for 
training at RAR cycle $m$, $\mathcal{S}\subset\mathcal{C}$ is the subset 
of $N_{\mathrm{c}}$ candidate points with the largest residual 
magnitudes, and $\mathcal{D}^{(m+1)}$ is the enriched collocation-point 
set obtained by adding $\mathcal{S}$ to $\mathcal{D}^{(m)}$. Thus, each 
RAR cycle augments the training set with points located in regions where 
the current approximation exhibits the largest governing-equation 
residuals.
The network is subsequently re-optimized using L-BFGS. This procedure 
is repeated for the prescribed number of refinement cycles. The complete 
procedure is summarized in Algorithm~\ref{alg:rar}.

\begin{algorithm}
	\caption{Hybrid Adam/L-BFGS training with Residual-Based Adaptive Refinement}
	\label{alg:rar}
	\begin{algorithmic}[1] 
		
		\State Initialize the collocation-point set $\mathcal{D}^{(0)}$;
		\State Initialize the network parameters $\theta$;
		\State Optimize the network using Adam;
		\State Optimize the network using L-BFGS;
		
		\For{$m=1,\ldots,N_{\mathrm{RAR}}$}
		\State Generate the candidate collocation-point set $\mathcal{C}$;
		\State Evaluate the physics residual over $\mathcal{C}$;
		\State Compute the residual magnitude $r_i$;
		\State Select the subset $\mathcal{S}\subset\mathcal{C}$ containing the $N_{\mathrm{c}}$ largest residuals;
		\State Update the collocation-point set $\mathcal{D}^{(m+1)}=\mathcal{D}^{(m)}\cup\mathcal{S}$;
		\State Optimize the network using L-BFGS;
		\EndFor
		
		\State \Return trained neural network $u_{\theta}(x,t)$;
		
	\end{algorithmic} 
\end{algorithm}

\subsection{Experimental Configuration}

All numerical experiments use the same network architecture, optimization protocol, and RAR strategy. The experiments are implemented in Python \cite{Python} using PyTorch \cite{Paszke2019} for automatic differentiation and neural-network optimization. Computations were performed using local CPU resources and, for the larger experiments, on an NVIDIA L40S GPU. The reference solution and implementation parameters are specified below to ensure reproducibility.

\subsubsection{Reference Solution}
\label{sec:referencesol}
All PINN-based models are evaluated against a high-resolution reference solution generated using the numerical solver from the official implementation of \citet{Raissi2019}. The solution is computed on a dense space--time grid with $N_x=2048$ spatial points and $N_t=101$ temporal snapshots, yielding 206,848 evaluation points.

The reference solution is used exclusively for post-training validation and error assessment and is not involved in training, adaptive refinement, or network optimization.

\subsubsection{Neural Network Configuration}

All models employ the Sinusoidal Representation Network (SIREN) architecture proposed by \citet{sitzmann2020siren}, whose periodic activations facilitate the representation of high-frequency solution components and sharp gradients. Following \citet{sitzmann2020siren}, the first-layer weights are initialized as
\begin{equation}
	W
	\sim
	\mathcal{U}
	\left(
	-\frac{1}{n_{\mathrm{in}}},
	\frac{1}{n_{\mathrm{in}}}
	\right),
	\label{eq:first_layer_init}
\end{equation}

while subsequent hidden layers use
\begin{equation}
	W
	\sim
	\mathcal{U}
	\left(
	-\frac{\sqrt{6/n_{\mathrm{in}}}}{\omega_0},
	\frac{\sqrt{6/n_{\mathrm{in}}}}{\omega_0}
	\right),
	\label{eq:hidden_layer_init}
\end{equation}

where $n_{\mathrm{in}}$ is the input dimension. Unless otherwise stated, $\omega_0=30$ is used throughout the experiments.

\begin{figure}
	\centering
	\begin{tikzpicture}[
		scale=0.95,
		transform shape,
		>=stealth,
		line width=0.8pt,
		layer/.style={
			draw,
			rounded corners=1pt,
			minimum width=0.95cm,
			minimum height=0.32cm,
			font=\scriptsize
		},
		dense layer/.style={
			draw,
			rounded corners=1pt,
			minimum width=0.6cm,
			minimum height=0.18cm,
			inner sep=0pt,
			text height=0.12cm,
			text depth=0.03cm,
			font=\tiny
		},
		dense arrow/.style={
			-{Stealth[length=0.7mm,width=0.7mm]}
		},
		input/.style={
			draw,
			circle,
			minimum size=0.52cm,
			font=\scriptsize
		},
		output/.style={
			draw,
			rounded corners=1pt,
			minimum width=0.95cm,
			minimum height=0.32cm,
			font=\scriptsize
		}
		]
		
		\node[font=\small\bfseries] at (2.2,5.0) {(a) PINN};
		
		\node[input] (pinn_in) at (2.2,4.25) {$(x,t)$};
		
		\node[layer] (pinn_l1) at (2.2,3.55) {$64$};
		\node[layer] (pinn_l2) at (2.2,2.85) {$64$};
		\node[layer] (pinn_l3) at (2.2,2.15) {$64$};
		\node[layer] (pinn_l4) at (2.2,1.45) {$64$};
		
		\node[output] (pinn_out) at (2.2,0.75) {$u_\theta$};
		
		\draw[->] (pinn_in) -- (pinn_l1);
		\draw[->] (pinn_l1) -- (pinn_l2);
		\draw[->] (pinn_l2) -- (pinn_l3);
		\draw[->] (pinn_l3) -- (pinn_l4);
		\draw[->] (pinn_l4) -- (pinn_out);

		\node[font=\small\bfseries] at (7.2,5.0) {(b) XPINN-SS};
		
		\node[input] (ss_in) at (7.2,4.25) {$(x,t)$};
		
		\node[layer] (ss_a1) at (6.15,3.50) {$64$};
		\node[layer] (ss_a2) at (6.15,2.85) {$64$};
		\node[layer] (ss_a3) at (6.15,2.20) {$64$};
		\node[layer] (ss_a4) at (6.15,1.55) {$64$};
		\node[output] (ss_ao) at (6.15,0.85) {$u_{\theta_1}$};
		
		\node[layer] (ss_b1) at (7.2,3.50) {$64$};
		\node[layer] (ss_b2) at (7.2,2.85) {$64$};
		\node[layer] (ss_b3) at (7.2,2.20) {$64$};
		\node[layer] (ss_b4) at (7.2,1.55) {$64$};
		\node[output] (ss_bo) at (7.2,0.85) {$u_{\theta_2}$};
		
		\node[layer] (ss_c1) at (8.25,3.50) {$64$};
		\node[layer] (ss_c2) at (8.25,2.85) {$64$};
		\node[layer] (ss_c3) at (8.25,2.20) {$64$};
		\node[layer] (ss_c4) at (8.25,1.55) {$64$};
		\node[output] (ss_co) at (8.25,0.85) {$u_{\theta_3}$};
		
		\draw[->] (ss_in) -- (ss_a1);
		\draw[->] (ss_in) -- (ss_b1);
		\draw[->] (ss_in) -- (ss_c1);
		
		\foreach \a/\b in {
			ss_a1/ss_a2,
			ss_a2/ss_a3,
			ss_a3/ss_a4,
			ss_b1/ss_b2,
			ss_b2/ss_b3,
			ss_b3/ss_b4,
			ss_c1/ss_c2,
			ss_c2/ss_c3,
			ss_c3/ss_c4}
		{
			\draw[->] (\a) -- (\b);
		}
		
		\draw[->] (ss_a4) -- (ss_ao);
		\draw[->] (ss_b4) -- (ss_bo);
		\draw[->] (ss_c4) -- (ss_co);
		
		\node[font=\small\bfseries] at (2.2,-0.1) {(c) XPINN-DS};
		
		\node[input] (ds_in) at (2.2,-0.85) {$(x,t)$};
		
		\node[layer] (ds_a1) at (1.15,-1.60) {$64$};
		\node[layer] (ds_a2) at (1.15,-2.25) {$64$};
		\node[layer] (ds_a3) at (1.15,-2.90) {$64$};
		\node[layer] (ds_a4) at (1.15,-3.55) {$64$};
		\node[output] (ds_ao) at (1.15,-4.25) {$u_{\theta_1}$};
		
		\node[dense layer] (ds_b1) at (2.2,-1.50) {$128$};
		\node[dense layer] (ds_b2) at (2.2,-1.85) {$128$};
		\node[dense layer] (ds_b3) at (2.2,-2.20) {$128$};
		\node[dense layer] (ds_b4) at (2.2,-2.55) {$128$};
		\node[dense layer] (ds_b5) at (2.2,-2.90) {$128$};
		\node[dense layer] (ds_b6) at (2.2,-3.25) {$128$};
		\node[dense layer] (ds_b7) at (2.2,-3.60) {$128$};
		\node[dense layer] (ds_b8) at (2.2,-3.95) {$128$};
		
		\node[output] (ds_bo) at (2.2,-4.35) {$u_{\theta_2}$};
		
		\node[layer] (ds_c1) at (3.25,-1.60) {$64$};
		\node[layer] (ds_c2) at (3.25,-2.25) {$64$};
		\node[layer] (ds_c3) at (3.25,-2.90) {$64$};
		\node[layer] (ds_c4) at (3.25,-3.55) {$64$};
		\node[output] (ds_co) at (3.25,-4.25) {$u_{\theta_3}$};
		
		\draw[->] (ds_in) -- (ds_a1);
		\draw[dense arrow] (ds_in) -- (ds_b1);   
		\draw[->] (ds_in) -- (ds_c1);
		
		\foreach \a/\b in {
			ds_b1/ds_b2,
			ds_b2/ds_b3,
			ds_b3/ds_b4,
			ds_b4/ds_b5,
			ds_b5/ds_b6,
			ds_b6/ds_b7,
			ds_b7/ds_b8}
		{
			\draw[dense arrow] (\a) -- (\b);
		}
		
		\foreach \a/\b in {
			ds_a1/ds_a2,
			ds_a2/ds_a3,
			ds_a3/ds_a4,
			ds_c1/ds_c2,
			ds_c2/ds_c3,
			ds_c3/ds_c4}
		{
			\draw[->] (\a) -- (\b);
		}
		
		\draw[->] (ds_a4) -- (ds_ao);
		\draw[dense arrow] (ds_b8) -- (ds_bo);   
		\draw[->] (ds_c4) -- (ds_co);
		
		\node[font=\small\bfseries] at (7.2,-0.1) {(d) XPINN-OV};
		
		\node[input] (ov_in) at (7.2,-0.85) {$(x,t)$};
		
		\node[layer] (ov_a1) at (6.15,-1.60) {$64$};
		\node[layer] (ov_a2) at (6.15,-2.25) {$64$};
		\node[layer] (ov_a3) at (6.15,-2.90) {$64$};
		\node[layer] (ov_a4) at (6.15,-3.55) {$64$};
		\node[output] (ov_ao) at (6.15,-4.25) {$u_{\theta_1}$};
		
		\node[layer] (ov_b1) at (7.2,-1.60) {$64$};
		\node[layer] (ov_b2) at (7.2,-2.25) {$64$};
		\node[layer] (ov_b3) at (7.2,-2.90) {$64$};
		\node[layer] (ov_b4) at (7.2,-3.55) {$64$};
		\node[output] (ov_bo) at (7.2,-4.25) {$u_{\theta_2}$};
		
		\node[layer] (ov_c1) at (8.25,-1.60) {$64$};
		\node[layer] (ov_c2) at (8.25,-2.25) {$64$};
		\node[layer] (ov_c3) at (8.25,-2.90) {$64$};
		\node[layer] (ov_c4) at (8.25,-3.55) {$64$};
		\node[output] (ov_co) at (8.25,-4.25) {$u_{\theta_3}$};
		
		\draw[->] (ov_in) -- (ov_a1);
		\draw[->] (ov_in) -- (ov_b1);
		\draw[->] (ov_in) -- (ov_c1);
		
		\foreach \a/\b in {
			ov_a1/ov_a2,
			ov_a2/ov_a3,
			ov_a3/ov_a4,
			ov_b1/ov_b2,
			ov_b2/ov_b3,
			ov_b3/ov_b4,
			ov_c1/ov_c2,
			ov_c2/ov_c3,
			ov_c3/ov_c4}
		{
			\draw[->] (\a) -- (\b);
		}
		
		\draw[->] (ov_a4) -- (ov_ao);
		\draw[->] (ov_b4) -- (ov_bo);
		\draw[->] (ov_c4) -- (ov_co);
		
	\end{tikzpicture}
	
	\caption{
		Neural-network configurations considered in this study:
		(a) PINN, (b) XPINN-SS, (c) XPINN-DS, and (d) XPINN-OV.
	}
	
	\label{fig:network_architectures}
\end{figure}
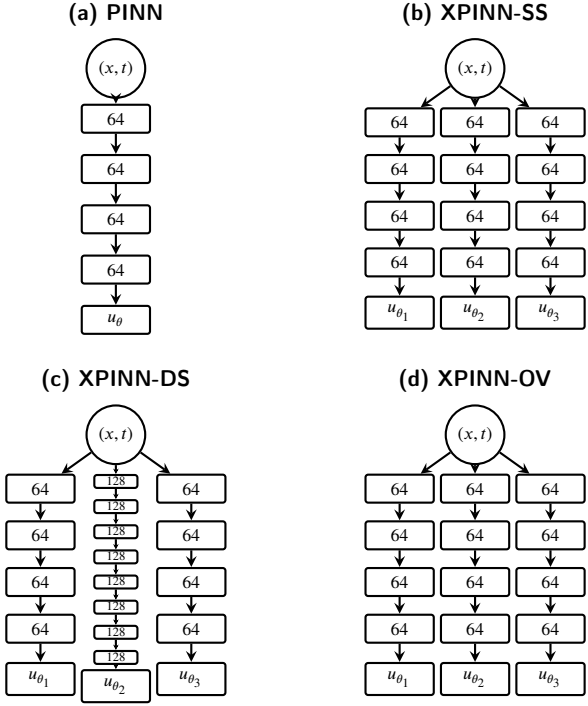

The network configurations considered in this study are illustrated in
Fig.~\ref{fig:network_architectures}. The corresponding architectural
parameters are summarized in Table~\ref{tab:architectures}.

\begin{table}[t]
	\centering
	\caption{Neural network architectures considered in this study.}
	\label{tab:architectures}
	\begin{tabular}{lccc}
		\hline
		Model & Subnetworks & Hidden layers & Neurons \\
		\hline
		PINN      & 1 & 4 & 64 \\
		XPINN-SS     & 3 & 4 & 64 \\
		XPINN-DS  & 3 & 4 / 8 / 4 & 64 / 128 / 64 \\
		XPINN-OV  & 3 & 4 & 64 \\
		\hline
	\end{tabular}
\end{table}

As summarized in Table~\ref{tab:architectures}, the PINN uses a single network, while XPINN-SS and XPINN-OV employ three subnetworks with four hidden layers and 64 neurons per layer. XPINN-DS increases the capacity of the central subnetwork to eight layers with 128 neurons per layer, while retaining the lateral architectures. XPINN-OV uses the same architecture as XPINN-SS but introduces overlapping subdomains.

\subsubsection{Domain Decomposition Configuration}

The computational domain $\Omega=[-1,1]$ was partitioned into three spatial subdomains, with interfaces located at
\begin{equation}
	x_L^\Gamma=-0.1,
	\qquad
	x_R^\Gamma=0.1.
\end{equation}

For the non-overlapping XPINN-SS and XPINN-DS configurations, the partition was defined as
\begin{equation}
	\begin{aligned}
		\Omega_1 &= [-1,-0.1), \\
		\Omega_2 &= [-0.1,0.1], \\
		\Omega_3 &= (0.1,1].
	\end{aligned}
\end{equation}

The central subdomain contains the shock formation region, while the lateral subdomains cover the smoother portions of the solution.

\begin{figure}
	\centering
	\begin{tikzpicture}[
		scale=0.80,
		transform shape,
		x=1.0cm,
		y=1.0cm,
		>=stealth,
		line width=0.8pt
		]
		
		\def\xmin{0}
		\def\xone{3}
		\def\xtwo{6}
		\def\xmax{9}
		\def\d{0.5}
		
		
		\fill[
		pattern=north east lines,
		pattern color=gray
		]
		(\xone-\d,-0.28)
		rectangle
		(\xone+\d,0.28);
		
		\draw[
		thick
		]
		(\xone-\d,-0.28)
		rectangle
		(\xone+\d,0.28);
		
		\fill[
		pattern=north east lines,
		pattern color=gray
		]
		(\xtwo-\d,-0.28)
		rectangle
		(\xtwo+\d,0.28);
		
		\draw[
		thick
		]
		(\xtwo-\d,-0.28)
		rectangle
		(\xtwo+\d,0.28);
		
		
		\draw[thick]
		(\xmin,0) -- (\xmax,0);
		
		
		\draw[thick]
		(\xmin,-0.12) -- (\xmin,0.12);
		
		\draw[thick]
		(\xmax,-0.12) -- (\xmax,0.12);
		
		
		\draw[dashed]
		(\xone,-0.65) -- (\xone,0.65);
		
		\draw[dashed]
		(\xtwo,-0.65) -- (\xtwo,0.65);
		
		
		\node[above=0.22cm] at (1.5,0)
		{$\Omega_1$};
		
		\node[above=0.22cm] at (4.5,0)
		{$\Omega_2$};
		
		\node[above=0.22cm] at (7.5,0)
		{$\Omega_3$};
		
		
		\node[below=0.25cm] at (\xmin,0)
		{$x_L=-1$};
		
		\node[below=0.25cm] at (\xmax,0)
		{$x_R=1$};
		
		
		\node[below=0.25cm] at (\xone,0)
		{$x_L^\Gamma=-0.1$};
		
		\node[below=0.25cm] at (\xtwo,0)
		{$x_R^\Gamma=0.1$};

		
		\draw[<->]
		(\xone-\d,-0.95) --
		(\xone+\d,-0.95);
		
		\node[below=0.02cm] at (\xone,-0.95)
		{$2\delta$};
		
		\draw[<->]
		(\xtwo-\d,-0.95) --
		(\xtwo+\d,-0.95);
		
		\node[below=0.02cm] at (\xtwo,-0.95)
		{$2\delta$};
		
	\end{tikzpicture}
	
	\caption{
		Overlapping decomposition of $\Omega=[-1,1]$ into three subdomains
		with overlap width $2\delta$ around the nominal interfaces.
	}
	\label{fig:overlapping_decomposition}
\end{figure}
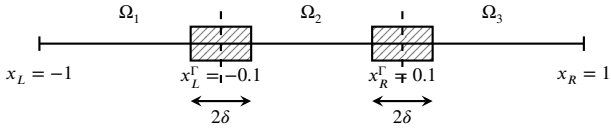

For the XPINN-OV formulation, neighboring subdomains were allowed to
overlap symmetrically around the nominal interfaces, as illustrated in
Fig.~\ref{fig:overlapping_decomposition}.

Let $\delta$ denote the
overlap extension from each nominal interface. The resulting
decomposition is defined as
\begin{equation}
	\begin{aligned}
		\Omega_1 &= [-1,\,x_L^\Gamma+\delta],\\
		\Omega_2 &= [x_L^\Gamma-\delta,\,x_R^\Gamma+\delta],\\
		\Omega_3 &= [x_R^\Gamma-\delta,\,1].
	\end{aligned}
\end{equation}

Three overlap extensions were investigated,
\begin{equation}
	\delta\in\{0.02,\;0.05,\;0.08\},
\end{equation}
These values were selected to assess the influence of overlap size on solution accuracy, conservative flux continuity, and information exchange between neighboring subnetworks.

\subsubsection{Training Configuration}

All numerical experiments were conducted using the same optimization protocol, summarized in Algorithm~\ref{alg:rar} and defined in Eq.~\ref{eq:eqprotocol}, to ensure a fair comparison among the PINN, XPINN-SS, XPINN-DS, and overlapping XPINN configurations. The latter were denoted as XPINN-OV0.02, XPINN-OV0.05, and XPINN-OV0.08, corresponding to overlap extensions of $\delta=0.02$, $0.05$, and $0.08$, respectively.
\begin{equation}
	\boxed{
		\begin{aligned}
			\mathrm{Adam}_{5000}
			&\;\rightarrow\;
			\mathrm{L\text{-}BFGS}_{3000}
			\;\rightarrow \\[-2pt]
			&\left[
			\mathrm{RAR}
			\;\rightarrow\;
			\mathrm{L\text{-}BFGS}_{3000}
			\right]_{10}
		\end{aligned}
	}
	\label{eq:eqprotocol}
\end{equation}

The initial collocation set consists of $200\times200=40,000$ points uniformly distributed over the entire space--time domain $\Omega\times I$. Thus, no preferential concentration of collocation points is introduced near the shock before the RAR procedure. For XPINN formulations, subsequent RAR enrichment is restricted to the central-region 	$\Omega_2$ subdomain.

During the Adam stage, the learning rate was initialized at $10^{-4}$ and exponentially decayed with $\gamma=0.999$ over 5000 epochs. Dynamic mini-batches of 20,000 collocation points were randomly sampled at each epoch, with gradient clipping applied using a maximum norm of 1.0.

The solution was subsequently refined using the L-BFGS optimizer with a Strong-Wolfe line search. Each L-BFGS stage used a maximum of 3000 iterations, a history size of 50, and convergence tolerances of $10^{-13}$ for parameter updates and $10^{-9}$ for the gradient norm.

RAR was performed for ten cycles. At each cycle, 200,000 candidate points were randomly generated and ranked according to the governing-equation residual, with the 500 points exhibiting the largest residuals added to the collocation set before the subsequent L-BFGS stage. 

The optimization hyperparameters are summarized in Table~\ref{tab:training_configuration}.

\begin{table}[T]
	\centering
	\caption{Training hyperparameters adopted in all numerical experiments.}
	\label{tab:training_configuration}
	\begin{tabular}{lc}
		\hline
		Parameter & Value \\
		\hline
		Numerical precision & Float64 \\
		Adam learning rate & $1\times10^{-4}$ \\
		Adam epochs & 5000 \\
		Learning-rate scheduler & Exponential ($\gamma=0.999$) \\
		Gradient clipping & 1.0 \\
		L-BFGS maximum iterations & 3000 \\
		L-BFGS history size & 50 \\
		Line search & Strong Wolfe \\
		Tolerance (parameter update) & $10^{-13}$ \\
		Tolerance (gradient norm) & $10^{-9}$ \\
		Initial collocation points & $200\times200 = 40\,000$ \\
		PDE mini-batch size & 20,000 \\
		Initial-condition points & 500 \\
		Boundary-condition points & 300 per boundary \\
		RAR candidate points & 200,000 \\
		RAR added points per cycle & 500 \\
		RAR refinement cycles & 10 \\
		\hline
	\end{tabular}
\end{table}

The loss function was defined as a weighted combination of the governing-equation residual, initial and boundary conditions, and, for domain-decomposed formulations, interface or overlap consistency terms. During Adam optimization, the PDE-residual weight was progressively increased according to

\begin{equation}
	w_{\mathrm{PDE}} =
	\begin{cases}
		10^{-4}, & 0 \leq n < 300, \\
		10^{-3}, & 300 \leq n < 1000, \\
		10^{-2}, & n \geq 1000.
	\end{cases}
	\label{eq:weight_schedule}
\end{equation}

where $n$ denotes the Adam epoch. The initial- and boundary-condition weight was fixed at $w_{\mathrm{ICBC}}=1$, while the interface and overlap terms were assigned unit weights, $w_{\mathrm{interface}}=1$ and $w_{\mathrm{overlap}}=1$, respectively.
\\

For the standard PINN, the total loss is given by Eq. (\ref{eq:total_loss_pinn}).
For XPINN, the loss additionally includes the interface consistency term,
\begin{equation}
	\begin{aligned}
		\mathcal{L}_{\mathrm{XPINN}}
		&=
		\sum_{k=1}^{N_s}
		\left[
		w_{\mathrm{PDE}}\mathcal{L}_{\mathrm{PDE}}^{(k)}
		+
		w_{\mathrm{ICBC}}
		\left(
		\mathcal{L}_{\mathrm{IC}}^{(k)}
		+
		\mathcal{L}_{\mathrm{BC}}^{(k)}
		\right)
		\right] \\
		&\quad+
		w_{\mathrm{interface}}\mathcal{L}_{\mathrm{interface}}.
	\end{aligned}
	\label{eq:xpinn_training_loss}
\end{equation}
where
\begin{equation}
	\mathcal{L}_{\mathrm{interface}}
	=
	\mathcal{L}_{u}^{\Gamma}
	+
	\mathcal{L}_{F}^{\Gamma}.
	\label{eq:xpinn_interface_loss}
\end{equation}

For XPINN-OV, the interface term is replaced by an overlap consistency loss,
\begin{equation}
	\begin{aligned}
		\mathcal{L}_{\mathrm{XPINN-OV}}
		&=
		\sum_{k=1}^{N_s}
		\left[
		w_{\mathrm{PDE}}\mathcal{L}_{\mathrm{PDE}}^{(k)}
		\right. \\
		&\qquad+
		\left.
		w_{\mathrm{ICBC}}
		\left(
		\mathcal{L}_{\mathrm{IC}}^{(k)}
		+
		\mathcal{L}_{\mathrm{BC}}^{(k)}
		\right)
		\right] \\
		&\quad+
		w_{\mathrm{overlap}}\mathcal{L}_{\mathrm{overlap}}.
	\end{aligned}
	\label{eq:xpinn_ov_training_loss}
\end{equation}
with
\begin{equation}
	\mathcal{L}_{\mathrm{overlap}}
	=
	\mathcal{L}_{u}^{\Omega}
	+
	\mathcal{L}_{F}^{\Omega}.
	\label{eq:xpinn_ov_overlap_loss}
\end{equation}

The final Adam weights, $w_{\mathrm{PDE}}=10^{-2}$, $w_{\mathrm{ICBC}}=1$, $w_{\mathrm{interface}}=1$, and $w_{\mathrm{overlap}}=1$, are retained during all subsequent L-BFGS and RAR refinement stages.

\section{Results}
\label{sec:results}

\subsection{RAR Convergence and Quantitative Performance}
\label{subsec:rar_quantitative}

The quantitative performance of all configurations is evaluated against
the reference solution described in Section~\ref{sec:referencesol}. Since the primary objective is accurate shock resolution, the best-performing checkpoint
for each configuration is selected according to the minimum relative
$L_2$ error in the central subdomain $\Omega_2=[-0.1,0.1]$. The
corresponding $L_\infty$ error and maximum PDE residual are evaluated at
the same RAR stage.

Table~\ref{tab:quantitative_results} summarizes the resulting
performance. The optimal RAR stage varies substantially across the
configurations, ranging from RAR~3 for XPINN-OV0.08 to RAR~9 for XPINN-DS.
For XPINN-DS, the subsequent RAR~10 stage produces a slight increase in
the solution error, indicating that additional refinement does not
necessarily improve the approximation.

\begin{table*}[t]
	\centering
	\caption{
		Quantitative performance of the considered configurations at their
		respective best RAR stages. The best stage is selected according to
		the minimum relative $L_2$ error in the central region
		$\Omega_2=[-0.1,0.1]$. Global and $\Omega_2$ solution errors are
		reported together with the maximum PDE residual in $\Omega_2$.
	}
	\label{tab:quantitative_results}
	\setlength{\tabcolsep}{7pt}
	\renewcommand{\arraystretch}{1.15}
	\begin{tabular}{l c c c c c c}
		\hline
		Configuration
		& Best RAR
		& \multicolumn{2}{c}{Global}
		& \multicolumn{3}{c}{$\Omega_2$}
		\\
		\cline{3-4}
		\cline{5-7}
		&
		&
		$L_2^{\mathrm{rel}}$
		& $L_\infty$
		&
		$L_2^{\mathrm{rel}}$
		& $L_\infty$
		& $\mathcal{R}_{\max}$
		\\
		\hline
		
		PINN
		& RAR 5
		& $1.029\times10^{-3}$
		& $1.031\times10^{-2}$
		& $2.704\times10^{-3}$
		& $1.031\times10^{-2}$
		& $2.469\times10^{-2}$
		\\
		
		XPINN-SS
		& RAR 6
		& $1.732\times10^{-4}$
		& $1.635\times10^{-3}$
		& $4.062\times10^{-4}$
		& $1.635\times10^{-3}$
		& $1.254\times10^{-2}$
		\\
		
		XPINN-OV0.02
		& RAR 7
		& $1.350\times10^{-4}$
		& $1.148\times10^{-3}$
		& $2.840\times10^{-4}$
		& $1.148\times10^{-3}$
		& $1.018\times10^{-2}$
		\\
		
		XPINN-OV0.05
		& RAR 4
		& $1.927\times10^{-4}$
		& $1.926\times10^{-3}$
		& $4.810\times10^{-4}$
		& $1.926\times10^{-3}$
		& $1.849\times10^{-2}$
		\\
		
		XPINN-OV0.08
		& RAR 3
		& $2.507\times10^{-3}$
		& $3.169\times10^{-2}$
		& $6.606\times10^{-3}$
		& $3.169\times10^{-2}$
		& $3.936\times10^{-2}$
		\\
		
		XPINN-DS
		& RAR 9
		& $5.480\times10^{-5}$
		& $2.810\times10^{-4}$
		& $7.180\times10^{-5}$
		& $2.810\times10^{-4}$
		& $3.086\times10^{-2}$
		\\
		
		\hline
	\end{tabular}
\end{table*}

Most XPINN configurations substantially outperform the standard PINN in
the central subdomain. XPINN-SS reduces the central relative $L_2$ error
by approximately $85\%$, while XPINN-OV0.02 provides a further reduction
of approximately $30\%$ relative to XPINN-SS. In contrast, increasing the
overlap width does not produce a monotonic improvement: XPINN-OV0.05
performs worse than XPINN-OV0.02, and XPINN-OV0.08 exhibits a central
relative $L_2$ error approximately $144\%$ larger than that of the
standard PINN.

XPINN-DS achieves the lowest solution errors among all configurations,
with a central relative $L_2$ error approximately $97\%$ lower than the
standard PINN. This result highlights the importance of representational
capacity in addition to spatial decomposition for resolving the
shock-dominated solution.

\begin{figure*}
	\centering
	\includegraphics[
	width=0.99\textwidth
	]{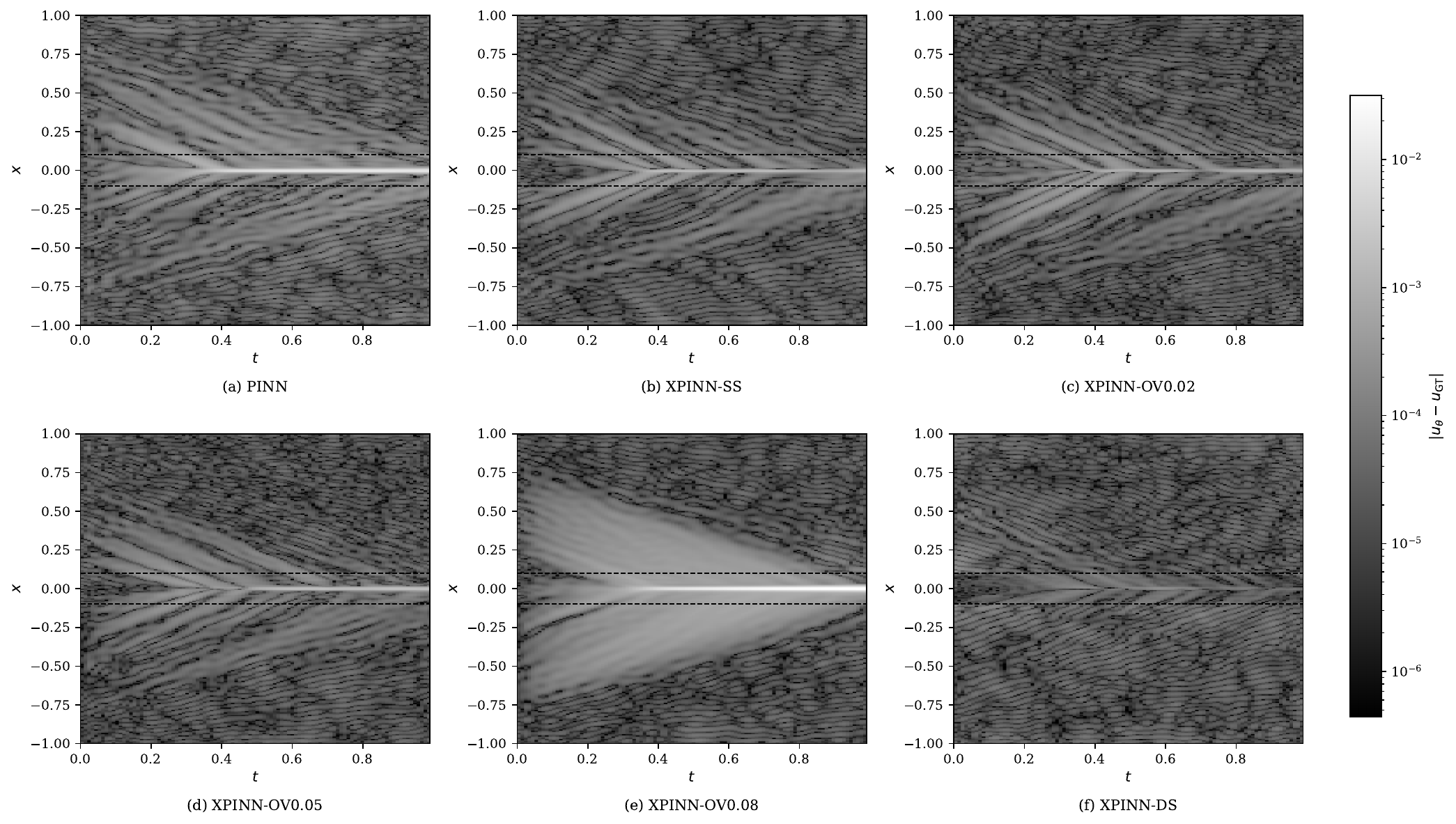}
	
	\caption{
		Absolute error,
		$|u_{\theta}-u_{\mathrm{GT}}|$,
		for the six considered configurations at their respective best
		RAR stages. The same logarithmic color scale is used for all
		panels to enable a direct comparison of the error magnitude.
		The dashed lines indicate the boundaries of the central region
		$\Omega_2=[-0.1,0.1]$.
	}
	\label{fig:absolute_error_comparison}
\end{figure*}

The spatial distribution of the absolute error for the six configurations
is shown in Figure~\ref{fig:absolute_error_comparison}. The common
logarithmic scale enables direct comparison of the error magnitude across
the different models.

\subsection{Training Dynamics and RAR Evolution}
\label{subsec:training_dynamics}

\begin{figure*}
	\centering

	\begin{subfigure}[b]{0.45\textwidth}
		\centering
		\includegraphics[width=\textwidth]{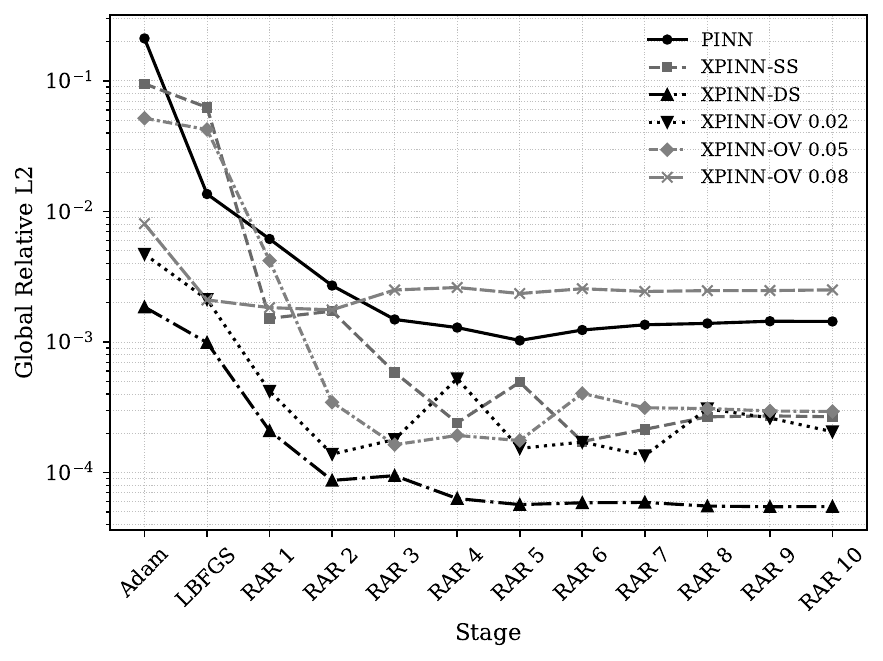}
		\caption{Global relative $L_2$ error.}
		\label{fig:global_l2}
	\end{subfigure}
	\hfill
	\begin{subfigure}[b]{0.45\textwidth}
		\centering
		\includegraphics[width=\textwidth]{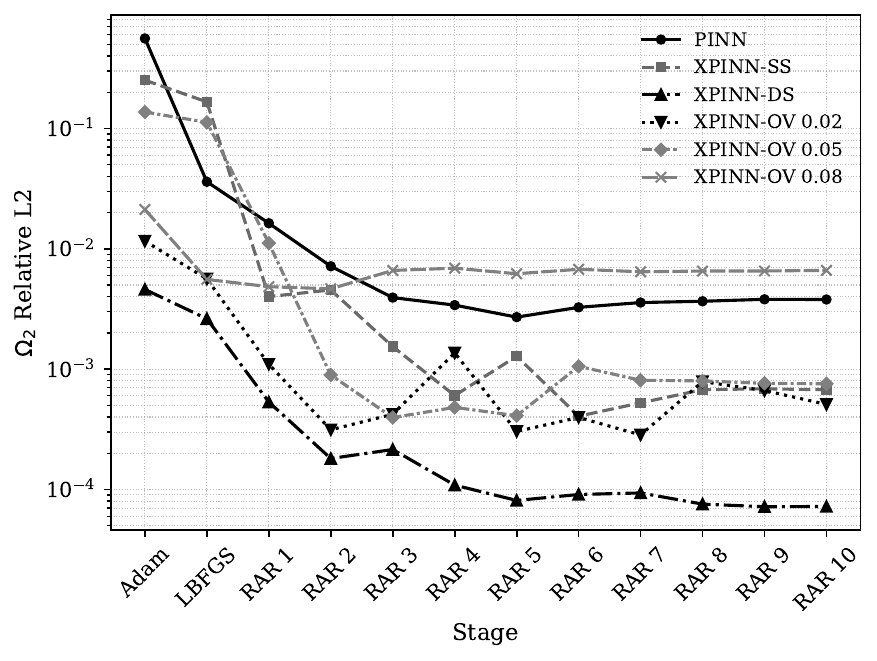}
		\caption{Relative $L_2$ error in $\Omega_2$.}
		\label{fig:shock_l2}
	\end{subfigure}
	
	\caption{
		Evolution of the relative $L_2$ error during the training procedure.
		Results are shown for the global domain and the central region
		$\Omega_2=[-0.1,0.1]$.
	}
	\label{fig:ev_l2}

\end{figure*}

To further examine the training behavior, the evolution of the solution
errors and PDE residuals is analyzed following the training protocol
established above. Figure~\ref{fig:ev_l2} shows the evolution of the global and
central-region relative $L_2$ errors during the training procedure.
During Adam, the configurations exhibit distinct accuracy levels, while
L-BFGS substantially reduces the errors and preserves differences between
the configurations.

The largest improvements occur during the first RAR cycles, particularly
RAR~1 and RAR~2. This behavior is consistent with the role of residual-
based refinement in concentrating additional collocation points near
regions that are initially under-resolved. For the present
shock-dominated problem, this provides additional training information
near the steep solution gradients, resulting in a rapid improvement in
the central region.

\begin{figure*}
	\centering	
	
	\begin{subfigure}[b]{0.45\textwidth}
		\centering
		\includegraphics[width=\textwidth]{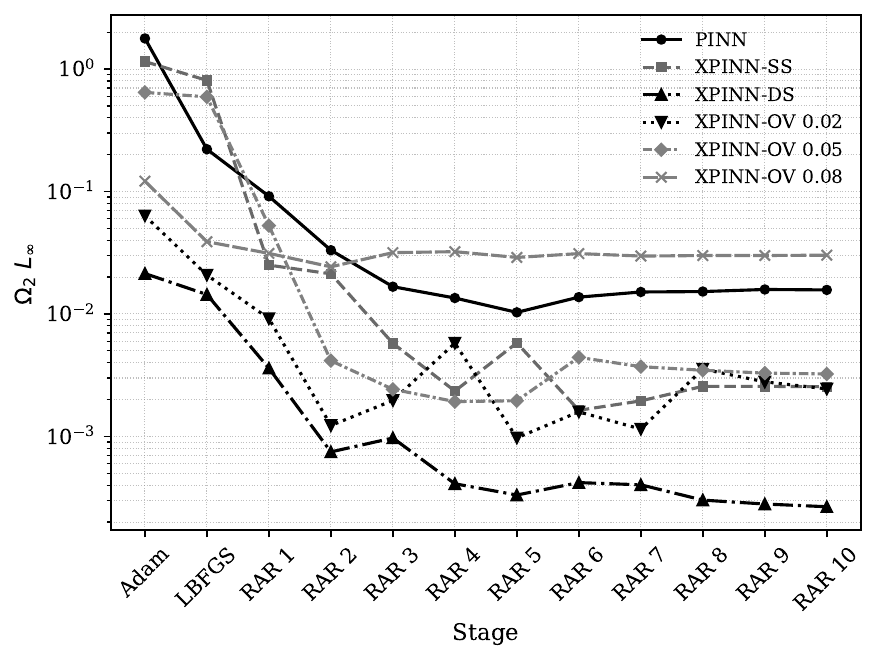}
		\caption{Central-region $L_\infty$ error.}
		\label{fig:shock_linf}
	\end{subfigure}
	\hfill
	\begin{subfigure}[b]{0.45\textwidth}
		\centering
		\includegraphics[width=\textwidth]{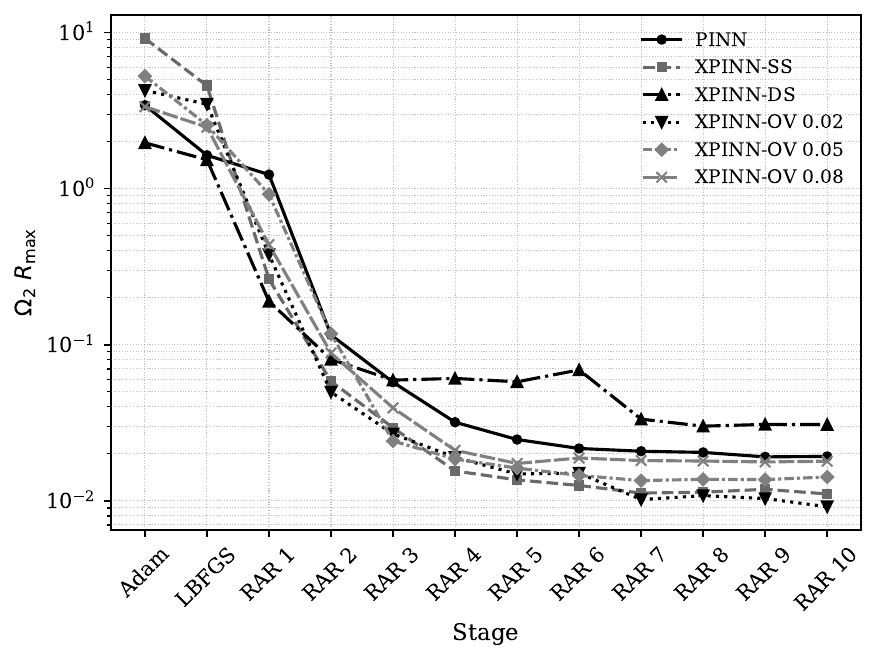}
		\caption{Maximum PDE residual $\mathcal{R}_{\max}$ in $\Omega_2$.}
		\label{fig:shock_maxres}
	\end{subfigure}
	
	\caption{
		Evolution of the central-region $L_\infty$ error and maximum PDE
		residual during the training procedure. Results are shown for
		$\Omega_2=[-0.1,0.1]$.
	}
	\label{fig:ev_accPDE}

\end{figure*}

After RAR~2, the error evolution becomes increasingly non-monotonic, with
subsequent cycles producing alternating improvements and degradations.
This indicates that additional point enrichment does not necessarily
translate into further global accuracy once the dominant shock region has
been sufficiently sampled. The $L_\infty$ error and maximum PDE residual
in Figure~\ref{fig:ev_accPDE} exhibit a similar trend, with strong
reductions during the initial refinement followed by smaller fluctuations.
Thus, later RAR cycles appear to provide progressively smaller local
corrections rather than systematic improvements.

The evolution of the overlapping configurations further shows that
increasing the overlap width does not systematically improve the
solution. Within the configurations considered, the smaller overlap
provides a more favorable evolution than the larger overlap, indicating
that the effectiveness of overlapping decomposition depends on the
selected overlap size rather than on the amount of shared information
alone. The double-shock configuration, in turn, maintains comparatively
stable errors during the later RAR cycles, suggesting that increased
representational capacity can complement adaptive collocation in
resolving the localized shock structure. 

\subsection{Effect of Overlap Width on Optimization Behavior} \label{subsec:overlap_behavior} 

\begin{figure*}
	\centering
	
	\begin{subfigure}[b]{0.32\textwidth}
		\centering
		\includegraphics[width=\textwidth]{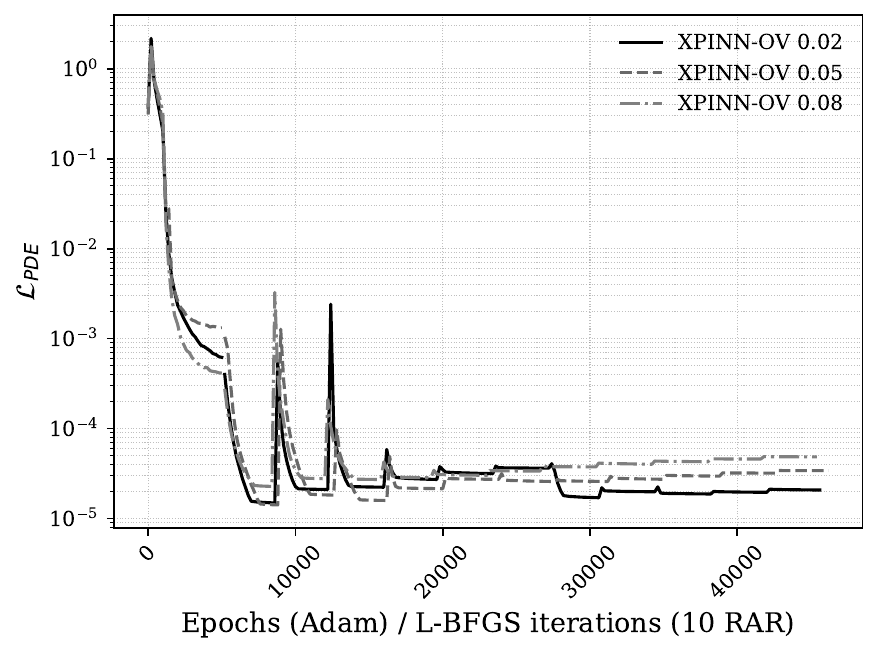}
		\caption{PDE loss}
		\label{fig:overlap_pde_loss}
	\end{subfigure}
	\hfill
	\begin{subfigure}[b]{0.32\textwidth}
		\centering
		\includegraphics[width=\textwidth]{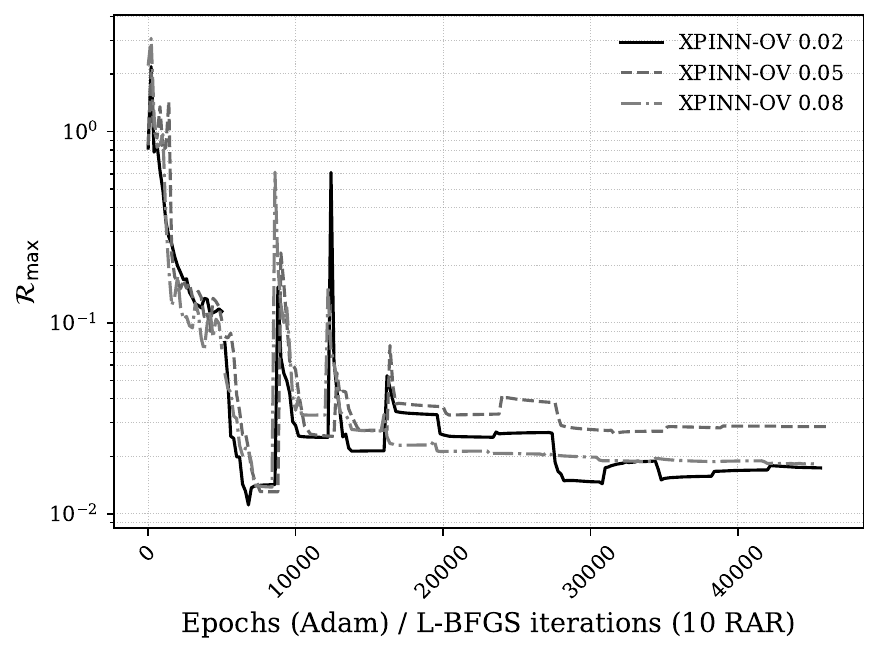}
		\caption{Maximum PDE residual $\mathcal{R}_{\max}$}
		\label{fig:overlap_max_residual}
	\end{subfigure}
	\hfill
	\begin{subfigure}[b]{0.32\textwidth}
		\centering
		\includegraphics[width=\textwidth]{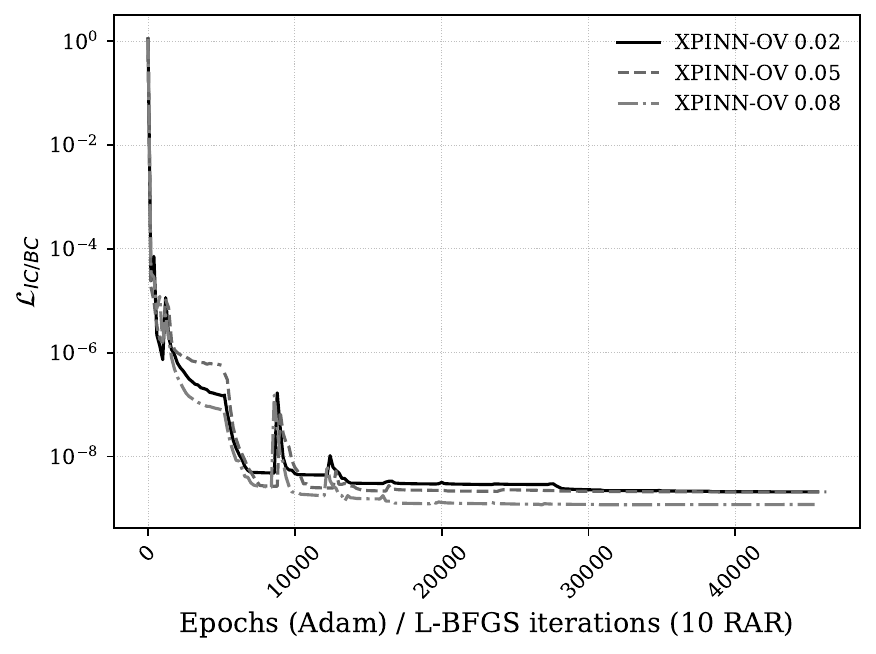}
		\caption{IC/BC loss}
		\label{fig:overlap_icbc_loss}
	\end{subfigure}
	
	\vspace{0.2cm}
	
	\begin{subfigure}[b]{0.32\textwidth}
		\centering
		\includegraphics[width=\textwidth]{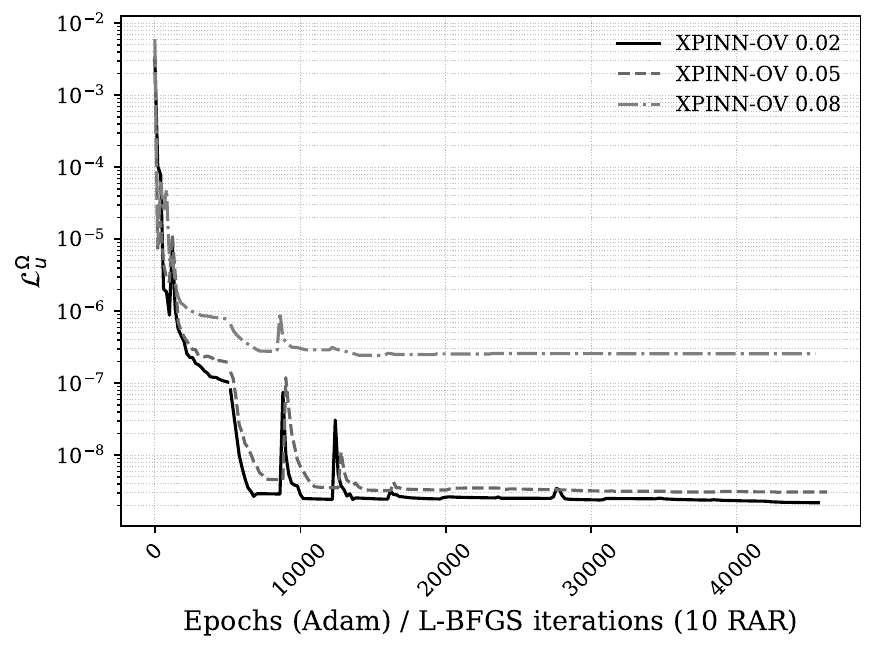}
		\caption{$u$-continuity loss}
		\label{fig:overlap_u_loss}
	\end{subfigure}
	\hspace{0.5cm}
	\begin{subfigure}[b]{0.32\textwidth}
		\centering
		\includegraphics[width=\textwidth]{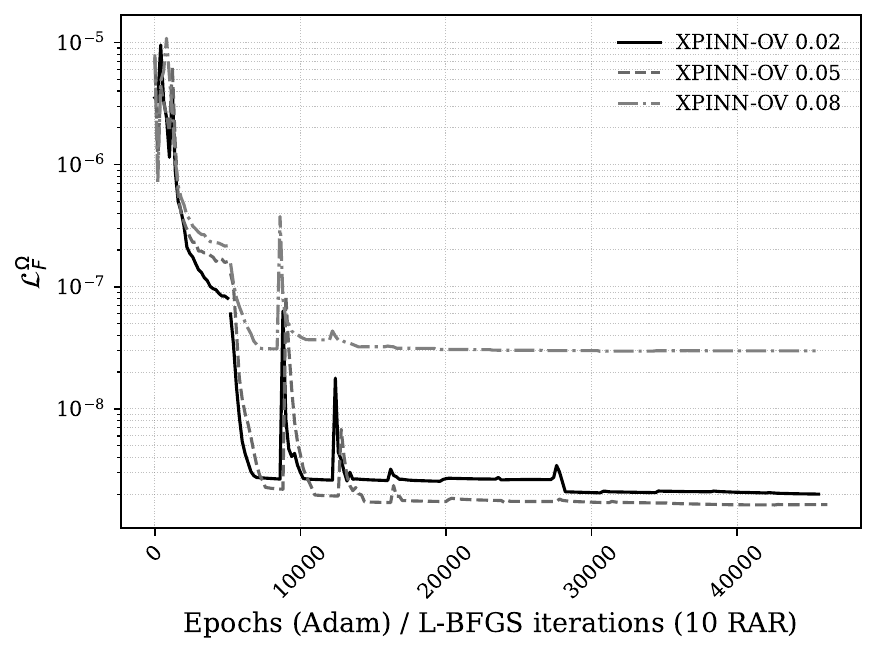}
		\caption{Flux-continuity loss}
		\label{fig:overlap_flux_loss}
	\end{subfigure}
	
	\caption{
		Evolution of the principal optimization quantities for the
		overlapping XPINN configurations with different overlap widths.
		The comparison includes the PDE loss, maximum PDE residual,
		initial and boundary-condition loss, and the interface losses
		associated with solution and flux continuity.
	}
	\label{fig:overlap_optimization}
\end{figure*}

The optimization dynamics shown in Figure~\ref{fig:overlap_optimization}
provide further insight into the effect of the overlap width. The comparison
includes the PDE loss, maximum PDE residual, initial and boundary-condition
losses, and the interface losses associated with solution and flux continuity
for $\delta=0.02$, $\delta=0.05$, and $\delta=0.08$.

During the initial Adam stage, all configurations exhibit a rapid reduction
in the PDE and IC/BC losses, followed by a slower decrease during the later
training stages. The maximum residual remains more sensitive to localized
regions, particularly near the shock, indicating that a reduction in the
global PDE loss does not necessarily correspond to an equivalent reduction
in the largest local residual. The adaptive refinement stages introduce
visible perturbations in the loss trajectories as new collocation points are
added, after which the subnetworks progressively adapt to the refined point
distribution.

The effect of the overlap width is most evident in the interface quantities.
The narrowest overlap, $\delta=0.02$, generally maintains lower
solution- and flux-continuity losses during the later stages than the wider
overlaps. Its maximum residual also exhibits a more favorable late-stage
evolution, whereas the wider configurations do not consistently recover to
lower levels after refinement. The IC/BC losses remain comparatively similar
among the three configurations, suggesting that the differences associated
with the overlap width are primarily related to the interaction between PDE
satisfaction and interface constraints rather than to the enforcement of
the prescribed conditions. These observations support the quantitative
results in Section~\ref{subsec:rar_quantitative}, where the narrow overlap
provided the most favorable overlapping configuration.

\section{Discussion}
\label{sec:discussion}

The results indicate that the performance of domain-decomposition PINNs
for the present shock-dominated problem depends on the interaction between
spatial decomposition, overlap, network capacity, and adaptive refinement,
rather than on any of these components individually. In particular, the
results show that domain decomposition can substantially improve shock
resolution, but that the effectiveness of the decomposition depends
strongly on how the subnetworks are coupled and represented.

\subsection{Domain Decomposition and Network Capacity}
\label{subsec:discussion_decomposition}

The improvement obtained by XPINN-SS over the standard PINN indicates that
spatial decomposition can facilitate the approximation of the localized
shock structure. By assigning separate subnetworks to different spatial
regions, the representation is localized and the central subnetwork can
focus on the region containing the steepest solution gradients. However,
the comparison among the XPINN configurations also shows that
decomposition alone is not sufficient to ensure improved accuracy.

The performance of XPINN-DS is particularly informative in this regard.
Its substantially lower solution error suggests that representational
capacity plays an important role in resolving the localized structure.
Thus, domain decomposition and network capacity appear to provide
complementary benefits: decomposition localizes the approximation,
whereas additional capacity increases the ability of the network to
represent the resulting solution features. This also indicates that
comparisons between domain-decomposition strategies should account for
the effective capacity of the resulting model rather than attributing
all improvements exclusively to the decomposition itself.

\subsection{Effect of Overlap and Optimization Behavior}
\label{subsec:discussion_overlap}

The overlapping configurations exhibit a clearly non-monotonic dependence
on the overlap width. The smallest overlap considered provides the best
performance among the overlapping cases, whereas increasing the overlap
does not lead to a corresponding improvement and eventually results in a
substantial degradation of accuracy. This suggests that a larger shared
region should not be interpreted simply as stronger or more effective
communication between neighboring subnetworks.

The optimization trajectories provide a possible explanation for this
behavior. The narrow-overlap configuration generally maintains lower
solution- and flux-continuity losses during the later stages of training,
together with a more favorable maximum-residual evolution. A larger
overlap therefore appears to introduce additional coupling requirements
without providing a proportional benefit for the solution approximation.
For the present problem, a relatively small overlap seems sufficient to
establish the required interaction between neighboring subnetworks while
avoiding unnecessary optimization complexity.

This interpretation should nevertheless be considered qualitative. The
present study does not directly quantify optimization stiffness through,
for example, Hessian spectra, condition numbers, or gradient statistics.
Similarly, spectral bias is used only as a possible interpretation of the
slower reduction of the remaining errors. The observed training behavior
is therefore consistent with these broader optimization phenomena, but it
does not isolate or establish their individual contributions.

\subsection{RAR and Implications of the Results}
\label{subsec:discussion_rar}

The RAR results further indicate that adaptive refinement is most effective
during the early stages of training. Once the dominant shock region has
been sufficiently enriched, additional refinement does not guarantee a
monotonic reduction in the solution error. This suggests that the value of
RAR depends not only on the number of added points but also on whether the
new points provide information that remains relevant to the unresolved
features of the solution.

More generally, the results suggest that the different components of the
training strategy address complementary aspects of the approximation
problem. RAR controls where additional physics constraints are introduced,
domain decomposition controls how the solution is spatially represented,
and network capacity controls the complexity of the functions that can be
represented by the subnetworks. The observed performance therefore cannot
be reduced to a single factor such as the number of collocation points or
the amount of overlap.

The conclusions should, however, be considered within the scope of the
present benchmark. The study is restricted to the one-dimensional viscous
Burgers equation and a limited set of overlap widths and network
configurations. A systematic variation of network capacity and overlap,
together with direct measurements of optimization conditioning and
spectral properties, would be required to establish the mechanisms
underlying the observed behavior more rigorously. Extending the analysis
to more complex shock-dominated and compressible-flow problems would
provide a further test of whether the observed trends generalize beyond
the present benchmark.

\section{Conclusions}
\label{sec:conclusions}

This study investigated the performance of domain-decomposition
Physics-Informed Neural Networks for the approximation of a shock-dominated
solution of the one-dimensional viscous Burgers equation. In particular,
the effects of spatial decomposition, overlap width, network capacity, and
residual-based adaptive refinement were evaluated using a common training
and evaluation framework.

The results show that spatial decomposition can substantially improve the
representation of the localized shock structure compared with a standard
PINN. However, the benefit of overlapping decomposition is not monotonic
with overlap width. Among the overlapping configurations considered, a
small overlap provided the most favorable performance, whereas a larger
overlap resulted in a substantial degradation. The double-shock
configuration achieved the highest solution accuracy, highlighting the
importance of representational capacity in addition to spatial
decomposition. These results indicate that decomposition, overlap, and
network capacity should be considered as coupled design choices rather
than independent sources of improvement.

RAR was most effective during the early
refinement stages, while later refinement produced smaller and
non-monotonic changes in the solution accuracy. The optimization analysis
also showed that overlap width primarily affected the interface-related
losses and the subsequent training dynamics. Overall, the results suggest
that effective shock resolution in domain-decomposition PINNs depends on
balancing spatial localization, representational capacity, interface
coupling, and adaptive point placement. Since the analysis is based on a
one-dimensional Burgers benchmark, further studies involving systematic
capacity and overlap variations, direct measurements of optimization
conditioning, and more complex shock-dominated or compressible-flow
problems are needed to assess the generality of these findings.

\section*{Acknowledgements}
This research was supported by resources supplied by the Center for Scientific Computing (NCC/GridUNESP) of the São Paulo State University (UNESP)

\printcredits

\bibliographystyle{model1-num-names-doi}

\bibliography{cas-refs}


%
%

\end{document}